%$\magnification=\magstep1
%\baselineskip=.33truein
%\font\bigtenrm=cmr10 scaled\magstep5
%\line{\hfill First Draft October 7, 2002}
%\bigskip
\centerline{\bf{Intelligent Life in Cosmology}}

\medskip

\medskip
\centerline {Frank J. Tipler}
\centerline {Department of Mathematics and
Department of Physics}
\centerline {Tulane University}
\centerline {New Orleans, Louisiana 70118 USA}
\medskip

\bigskip

\centerline {Abstract}

\bigskip

I shall present three arguments for the proposition that intelligent life is very rare in the universe.  First, I shall summarize the consensus opinion of the founders of the Modern Synthesis (Simpson, Dobzhanski, and Mayr) that the evolution of intelligent life is exceedingly improbable. Second, I shall develop the Fermi Paradox: if they existed, theyÕd be here.  Third, I shall show that if intelligent life were too common, it would use up all available resources and die out.   But I shall show that the quantum mechanical principle of unitarity (actually a form of teleology!) requires intelligent life to survive to the end of time.  Finally, I shall argue that, if the universe is indeed accelerating, then survival to the end of time requires that intelligent life, though rare, to have evolved several times in the visible universe.   I shall argue that the acceleration is a consequence of the excess of matter over antimatter in the universe.  I shall suggest experiments to test these claims.

\bigskip
\centerline{\bf{1. Introduction}} 
\bigskip

Martin Rees is fond of arguing, Òabsence of evidence is not evidence of absence.  How could anyone disagree?  But on the question of the existence of extraterrestrial intelligent life, we have an undeniable fact: they aren't here.  That is, extraterrestrial intelligent beings are not obviously present on our planet, or in our solar system.  I think even Martin will agree with this!  But I claim this fact allows us to conclude that extraterrestrial intelligence (ETI) is absence from our Galaxy and from the Local Group of galaxies.  In other words, if they existed, theyÕd be here!

This argument has often been called the Fermi Paradox.  I think it is analogous to Olbers' Paradox in cosmology, which uses an equally obvious fact, known to all of us --- the fact that the sky is dark at night --- to conclude that the universe must have evolved to its present state.  The universe cannot have been the same as it appears now for all eternity.  I shall outline in Section 2 the reasons that the absence of ETI on Earth allows us to conclude that they don't exist in our galactic neighborhood.  I have developed this argument is much more detail elsewhere, addressing all counter-arguments that have been proposed.  So I shall only outline my argument in Section 2.  I shall also only outline the evolutionary argument against ETI here.  Mayr, Dobzhanski, Simpson and Ayala have defended this position at length over the past 40 years, and I'm sure this argument is quite familiar to the readers of this journal.  What I want to develop in this paper is a new argument against the existence of ETI.

I shall call it the Limited Resources Argument.  It is related to the Fermi Paradox in that it assumes that an intelligent life form will inevitability expand off its planet of origin and once this expansion begins, it will never stop.  But if intelligent life were common in the cosmos, the expansion of technological civilization would use up resources so fast that intelligent life would die out.  If intelligent life is rare, the speed of light barrier will prevent life from using up the resources too fast.

The immediate reaction to this argument is, so what if intelligent life uses up the resources too fast and dies out?  Do we have any reason for believing that intelligent has some guarantee for survival that other species do not?  Most species that have evolved are now extinct, and have left no descendants.  Why should Homo sapiens be any different?  There is no evidence from evolutionary biology that intelligence should survive indefinitely.

But there is evidence from physics for the importance of intelligence life in cosmology.  Not of course in the current phase of universal history, but instead near the end of the universe.  
\bigskip\bigskip
\centerline{\bf{ II. Why Intelligent Life Must Be Rare}} 
\bigskip

\centerline{A.  The Improbable Evolution Argument}

\medskip
The argument against ETI that most readers of this journal will be familiar with goes back to Alfred Russell Wallace, and has more recently been defended by such major evolutionists as George Gaylord Simpson, Theodosius Dobzhanski, and Ernst Mayr.  These scientists point out that according to the Modern Synthesis, evolution has no knowledge of goals.  Instead, natural selection acts on random mutations, mutations which never appear with the intent of achieving a goal in the distant future.  There are an enormous number of evolutionary pathways, and so few of these lead to intelligent life, that it is unlikely intelligent life will appear more than once in the visible universe, which is the part of the universe within 13.7 billion light years.   The universe is observed to be 13.7 billion years old, and so we cannot see out a distance greater than 13.7 billion light years, the distance light could have traveled in that time.  (Actually, we can see out a bit further than 13.7 billion light years because of the expansion of the universe, but let me ignore this minor technicality.)  Even if we were to assume that all the matter and energy in the visible universe were in the form of Earthlike planets, there would be only (!) about $10^{28}$ Earthlike planets in the visible universe.  This number assumes that ``earthlike'' means only that the mass of the planet is greater than or equal to the mass of the Earth.  No assumption is made about the planet's star, atmosphere, or orbital radius.

	The well-known evolutionist Francisco Ayala has recently made this argument quantitative.  He estimates that the probability of an intelligent species evolving on an Earthlike planet upon which one-cell organisms have appeared is less than 10 to the minus one million power!  This number is so tiny that the evolution of intelligent life is exceedingly unlikely to have occurred even once.  Ayala's number is not contradicted by the fact that intelligent life exists on Earth.  It is just exceedingly improbable that it exists anywhere in the universe (at least if the universe is finite in spatial size, as I shall argue in Section IV that it is).  Ayala's number depends on the assumption that gene changes upon which natural selection operates are essentially random.  Evolution has no foresight.  Mayr has emphasized that intelligence on earth is limited to the chordate lineage, so, he argues that if the chordates never appeared on Earth, neither would intelligence.  But chordates first evolved more than half a billion years ago.  These animals did not know that they had to evolve so that Homo sapiens would eventually appear.  Natural selection can only operate during an animal's lifetime.  It cannot select a genome with the intent of using the genome a billion yeas later.

	There is an important caveat to this; a caveat first pointed out by Charles Darwin himself in the last pages of his book {\it The Variation of Animals and Plants under Domestication}.  Darwin noted that at the ultimate level of physics, the universe is deterministic.  This means that at the ultimate level, there are no random events.  In particular, the evolution of Homo sapiens was inevitable, determined by the initial state of the universe and the universeÕs initial conditions.  ``Random'' variation does not mean uncaused.  It just means unpredictable for human beings.  Therefore, at this ultimate physical level, Darwin claims that his own theory is only an approximation.  Darwin noted that the advance of science might enable us to obtain enough information to predict these ``random'' variations.  I shall argue below that this time has now come.
\bigskip
\centerline{B. If They Existed, They'd Be Here} 
\bigskip

The argument against the existence of extraterrestrial intelligent life that I have developed in most detail is sometimes called the Fermi Paradox: if they existed, they'd be here.  The force of this argument is not usually appreciated, because most people --- and even most scientists (! ---  tacitly assume that any alien civilization, no matter when they evolved or how long they have had advanced technology, will nevertheless have essentially the technology of the late 20th century.  The reason for this tacit assumption is the usual human weakness: we have an unfortunate habit of trying to impose our current human perspectives on the physical universe.  

But let consider the consequences of only slightly more advanced computer technology than we now have.  According to most computer experts, within a century or so we should have computer programs which have human level intelligence, computers which can run such programs and also make copies of themselves and the programs.  Imagine such a machine combined with our rocket technology into a space probe.  Such a space probe can reach the nearest star in 40,000 years.  Once there in the nearest star system, the probe could make several copies of itself, using the asteroid material which we now know is present in almost all star systems, sending these Ò daughterÓ probes to further star systems, where the process would be repeated.  Even with our rocket technology, every star system in the entire Galaxy would have a probe within 100 million years.  With a more advanced rocket technology, a rocket technology which is even today been experimented with, it should be possible to send a probe between the stars at 1/10 light speed.  With such a speed, probes would cover the entire galaxy within a few million years.  And all for the cost of a single probe!

Almost any motivation we can imagine would lead an intelligent species with the technology to launch that single probe.  Suppose for example, ET wants to contact other intelligent life forms.  Then rather than send out radio signals, they should send out that single probe.  With radio, one has to send out the signals to many stars, over many thousands of years.  (We would expect evolution to intelligence to require billions of years, as it did on Earth.)  But once the probe is launched, coverage of the entire galaxy is automatic.  Once in a target star system, the intelligent probe can contact any intelligent life forms that happen to have evolved on any planet in the system.  Or if no intelligent life is found, the probe can study the entire system and transmit the results back to Earth.  This on the spot investigation is obviously impossible if radio signals are sent out instead of a space probe. 

One might think an intelligent species would be reluctant to use probes because of the worry that these machines would eventually escape from the control of the original transmitting species.  But the same objection can be made to sending out radio signals.  It is impossible to predict what use a recipient species would make of the information in the signal.  Many scientists here on Earth have opposed the transmission of signals, fearing that hostile aliens may use the signals to home in on our planet.  The fear of losing control of the probes --- which, since these machines are rational beings, should be regarded as our mind children --- apply with equal force to our biological descendants.  ``No species now existing will transmit its unaltered likeness to a distant futurity'' was how Darwin put in the closing pages of {\it Origin of Species}.  We do not know whether they will be good or bad by our standards.  We do know that in the far future they won't be Homo sapiens.  

But in the long run, our descendants, whatever they look like, whether they are silicon machines or the more familiar DNA devices, must leave the Earth if they are to survive.  Within 6 billion years, the Sun's atmosphere will expand out and engulf the Earth, which will spiral into the Sun and be vaporized.  A similar fate is in store for any and all intelligent species that evolve on a water planet.  Making the reasonable Darwinian assumption that survival will be a central motivation of all intelligent species, all intelligent species will eventually develop space travel, leave their planet, and colonize their own star system.  The universe is 13.7 billion years old, and most stars and their planets are billions of years older than our own.  Thus, whatever the probability intelligent life evolves on an earthlike planet on which one-cell organisms appear, most intelligent species would be billions of years older than we are.  They should have left their mother planet billions of years ago.  Once they leave their planet, nothing can stop their expansion into interstellar space.  If they existed, they would be here.

\bigskip
\centerline{C. The Limited Resources Argument} 
\bigskip

Once an intelligent species begins its expansion into interstellar space, there is only the speed of light barrier to stop the expansion.  Furthermore, as Dyson has emphasized, intelligent life will eventually develop the ability to convert any form of matter into living matter and life support devices.  Given time, intelligent life can take apart no only asteroids, but also entire Jupiter-sized planets and even stars.  Thus a galaxy which has been invaded (infected?) by a space travelling intelligent life form will start to disappear.  This, by the way, is yet another argument for human uniqueness in the visible universe.  We have never observed galaxies in the process of controlled disintegration.  Intelligent life, in the long term, ought to appear as a horde of locusts, devouring all matter in its domain.  A galactic wide government cannot be set up to stop such behavior because of the speed of light barrier, but even if it could be set up, it would have no choice but to allow such behavior.  Survival requires the conversion of matter into energy.  Setting an ultimate limit to how much matter can be so converted would merely doom life to extinction.

However, the speed of light barrier, which prevents a galactic scale government from being set up to prevent life from devouring all matter, itself imposes a limitation on how fast life can use up resources.  The disc of our galaxy is some 100,000 light years across; we not use up the material resources of our galaxy in less than 100,000 years.  The Virgo cluster is some 60 million light years away.  We cannot use up the resources of the Virgo cluster in less than 60 million years.  If the universe were closed and decelerating, a single intelligent life form could not devour the entire universe until after the universe had begun to recollapse.  Actually the universe is currently accelerating.  If this acceleration were to continue forever at its present rate, our descendants could devour only the region currently within at most 10 billion light years.  This limit is imposed by the speed of light barrier modified by the universal acceleration.  

But the more intelligent life there is in the universe, the more planets upon which intelligent life independently evolves, the more rapidly resources will be used up.   When all the material resources are used up, intelligent life will die.  The more common intelligent life is in the universe, the more rapidly it will become extinct.  

Conversely, if intelligent life is quite rare --- a single intelligent species, if the universe were closed and always decelerating --- intelligent life would be forced by the laws of physics to use resources at just the right rate to survive to the very end of time.  And even more intelligent species could so survive if the universe were to have a period of acceleration in its expansion phase, as the universe is indeed observed to have.

But why should the universe adjust the number of intelligent species so that the descendants of the species would survive to the end of time?  As Darwin pointed out in the closing pages of {\it Origin of Species}, almost all species that have ever existed on Earth have died out, leaving no descendants.  Why should an intelligent life form have a survival probability utterly different from almost all other species?  I claim that intelligent life will survive until the end of time because the laws of physics require it. Or to put it another ways, because such survival is one of the goals of the universe.

\bigskip
\centerline{\bf{III. Unitarity is Teleology}} 
\bigskip

Teleology has been completely rejected by evolutionary biologists.  This rejection is unfortunate, because, teleology is alive and well in physics, under the name of Òunitarity.Ó  Unitarity is an absolutely central postulate of quantum mechanics, and it has many consequences.  One of these consequences is the CPT theorem, which implies that the g-factors of particles and antiparticles must be exactly equal.  This equality (for electrons and positrons) has been verified experimentally to 13 decimal places, the most precise experimental number we have.  Which is why very few physicists are willing to give up the postulate of unitarity!  Furthermore, unitarity is closely related to the law of conservation of energy, and a violation of unitarity has been shown to result usually in the gigantic creation of energy out of nothing.  One model (due to Leonard Susskin) of unitarity violation had the implication that whenever a microwave oven was turned on, so much energy was created that the Earth was blown apart.   So physicists are very reluctant to abandon unitarity. 

Unitarity is most often applied to what physicists call the S-matrix, which is the quantum mechanical linear operator that transforms any state in the ultimate past to a unique state in the ultimate future.  But unitarity more generally applies to the time evolution operator, a linear operator that carries the quantum state of the universe at any initial time uniquely into the quantum state of the universe at any chosen future time.  ÒUniquelyÓ is a key word.  It means that unitarity is the quantum mechanical version of determinism.   Contrary to what is generally thought, determinism is alive and well in quantum mechanics.  Determinism, however, applies to wave functions (quantum states) rather than to individual particles.  Alternatively, we can say that determinism applies to coherent collections of worlds rather than to individuals.  There is a sense, which I won't have room to discuss here, in which quantum mechanics is more deterministic than classical mechanics, and that Schr\"odinger derived his famous equation by requiring that classical mechanics in it most general expression (Hamilton-Jacobi theory) be deterministic. (See Tipler (2005) for the mathematical details.)

But the usual past-to-future determinism is not the fundamental meaning of unitarity.  What unitarity really means is that the inverse of the time evolution operator exists, and is easily computed from the time evolution operator itself by forming the time evolution operator's hermitian conjugate.  Any operator whose inverse is obtained in this manner is said to be a unitary operator.  But in the present context, the important point is that the inverse of the time evolution operator exists.  The inverse of any operator is an operator that undoes the effect of the original operator.  In the case of the time evolution operator, which generates past-to-future evolution, the inverse operator generates future-to-past evolution.  In other words, it carries future quantum states uniquely into past quantum states.  Therefore, unitarity tells us that any complete statement of usual past-to-future causation is mathematically equivalent to some complete statement of future-to-past causation.  In more traditional language, a complete list of all efficient causes is equivalent to some complete list of final causes.  Teleology is reborn!

Nevertheless, the Second Law of Thermodynamics says that the complexity of the universe at the microlevel is increasing with time.  This means that it will usually be the case that past-to-future causation will be the simpler explanation of the two causal languages.  But this will not always be the case.  We should always remember that for physical reality the two causation languages are mathematically equivalent.  It might occasionally be the case that we humans can understand where the evolution of the universe is taking us only by using future-to-past causation.  That is, we can understand what is happening now only by considering the ultimate goal of the universe.

To reject this possibility is a terrible mistake.  Humans naturally think in terms of past-to-future causation because our memories are designed (by the laws of physics) to work in this time direction.  But the universe is not similarly restricted.  It is a mistake to impose human limitations on the physical universe.  It was a terrible mistake to require that solar system mechanics look simple in a geocentric frame of reference.

Let me now use this future-to-past causation to show that biological evolution cannot be completely random.  I shall now argue that the laws of physics require intelligent life to evolve somewhere, and survive to the very end of time.

\bigskip
\centerline{\bf{IV. Why Intelligent Life MUST Exist in the Far Future}} 
\bigskip

	The necessity of intelligent life in the far future is an automatic consequence of the laws of physics, specifically quantum mechanics, general relativity, the Standard Model of particle physics, and most importantly, the Second Law of Thermodynamics.  I shall show that the mutual consistency of these laws requires three things.  First, the universe must be closed (the universe's spatial topology must be a three-sphere).  Second, life must survive to the very end of time.  Third, the knowledge possessed by life must increase to infinity as the end of time is approached.  I do not assume life survives to the end of time.  Life's survival follows from the laws of physics.  If the laws of physics be for us, who can be against us?

	But before I prove that the laws of physics require life to survive, let me first show that it is possible for life to survive.  To survive for infinite experiential time, life requires an unlimited supply of energy.  That is, the supply of available energy must diverge to infinity as the end of time is approached.  Nevertheless, conservation of energy requires the total energy of the universe to be constant.  In fact, Roger Penrose has shown that the total energy of any closed universe is ZERO!  The total energy is zero now, was zero in the past, and will be zero at all times in the future.  One might wonder how this is possible.  After all, we are now receiving energy from the Sun, we are using food energy as we read this, and we can extract energy from coal, oil, and uranium.  Energy, in other words, seems to be non-zero.

	However, the forms of energy just listed are not all the forms of energy in the universe.  There is also gravitational energy, which is negative.  So if we were to add all the positive forms of energy --- radiant energy, the stored energy in coal, oil, and uranium, and most importantly, the mass-energy of matter --- to the negative gravitational energy, the sum is zero.  This means that if we can make the gravitational energy even more negative, the positive energy, that is, the energy available for life, necessarily increases, even though the total energy in the universe stays zero.  The key property of energy that must always be kept in mind is that it transforms from one form to another. Once we realize that gravitational energy can transformed into available energy, we understand where life can obtain the unlimited source of available energy it needs for survival: life must make the total gravitational energy approach minus infinity.

	Life can do this only if the universe is closed, and collapses to zero size as the end of time is approached.  Conversely, if the universe is closed and collapses to zero size, then the total gravitational energy goes to minus infinity, since the gravitational energy of a system is inversely proportional the size of the system.  I have shown in my book (Tipler, 1994) that life can in fact extract unlimited available energy from the collapse of the universe.

	Now let me outline the proof of my three claims above.  I can give here only a bare outline.  For complete details, the reader is referred to my book (Tipler, 1994) and to papers ((Tipler et al, 2000), and (Tipler 2001)) on arXiv, the physics preprint database (available on the Internet at  http://arxiv.org/).  Black holes exist, but Hawking proved that were black holes to evaporate completely --- as they necessarily would if the universe were to expand forever --- the black holes would violate unitarity, the fundamental law of quantum mechanics which I described in the previous section.  Hence the universe must eventually stop expanding, collapse, and end in a final singularity.  If this final singularity were to be accompanied by event horizons, then the Bekenstein Bound (another law of quantum mechanics, basically the Heisenberg Uncertainty Principle expressed in the language of information theory) would have the following effect.  It would force that all the microstate information in the universe to go to zero as the universe approaches the final singularity.  But the microstate information going to zero would imply that the entropy of the universe would have to go to zero, and this would contradict the Second Law of Thermodynamics, which says that the entropy of the universe can never decrease.  But if event horizons do not exist, then the Bekenstein Bound allows the information in the microstates to diverge to infinity as the final singularity is approached.  Conversely, ONLY if event horizons do not exist can quantum mechanics (the Bekenstein Bound) be consistent with the Second Law of Thermodynamics.  Therefore, event horizons cannot exist, and by Seifert's Theorem (see (Tipler, 1994), p. 435) the non-existence of event horizons requires the universe to be spatially closed.  In Penrose's c-boundary construction (Tipler, 1994), (Hawking and Ellis, 1973), a singularity without event horizons is a single point.  I call such a final singularity the OMEGA POINT.  At a Windsor Castle conference, Martin Rees objected that many physicists (in particular, himself) do not accept Hawking's proof that unitarity would be violated were a black hole to evaporate to completion.  But most of the physicists who reject Hawking's argument nevertheless accept that there is nevertheless a Black Hole Information Problem: i.e., that we must explain how the information that falls into a black  hole gets out.  Many solutions to the Information Problem have been proposed but all of these solutions (except the one I shall advance) have one feature in common.  They all involve proposed new laws of physics.  My proposal --- that there are no event horizons at all, hence no black hole event horizons, so ALL information at all events are accessible to all observers in the far future --- does NOT involve new physical laws.  Only classical general relativity is used.  I use HawkingÕs unitarity argument only to infer the non-existence of event horizons.  If we resolve the Black Hole Information Problem by simply assuming the non-existence of event horizons, then I don't need to use either the Bekenstein Bound or the Second Law of Thermodynamics to infer the existence of the Omega Point, or spatial closure.   Resolving the Information Problem using known physics automatically yields no event horizons and spatial closure for the universe.

	If the universe were to evolve into an Omega Point type final singularity without life being present to guide its evolution, then the non-existence of event horizons would mean that the universe would be evolving into an infinitely improbable state.  Such an evolution would contradict the Second Law of Thermodynamics, which requires the universe to evolve from less probable to more probable states.  On the other hand, if life is present guiding the evolution of the universe into the final singularity, then the absence of event horizons is actually the MOST probable state, because the absence of event horizons is exactly what life requires in order to survive (details in my book (Tipler 1994)).   In other words, the validity of the Second Law of Thermodynamics REQUIRES life to be present all the way into the final singularity, and further, the Second Law requires life to guide the universe in such a way as to eliminate the event horizons.  Life is the only process consistent with known physical law capable of eliminating event horizons without the universe evolving into an infinitely improbable state.  Exactly how life eliminates the event horizons is described in my book (Tipler, 1994).  Roughly speaking, life nudges the universe so as to allow light to circumnavigate the universe first in one direction, and then another.  This is done repeatedly, an infinite number of times.  There are thus an INFINITE number of circumnavigations of light before the Omega Point is reached.  If we were to regard a single circumnavigation as a single tick of the Òlight clockÓ there would be an infinite amount of such time between now and the Omega Point.  An even more physical time would be the number of experiences which life has between now and the Omega Point.  This ``experiential time'' --- the time experienced by life in the far future --- is the most appropriate physical time to use near the Omega Point.  It is far more appropriate than the human based Òproper timeÓ we now use in our clocks.
\bigskip
\centerline{\bf{V.  Life in the Future of an Accelerating Universe}} 
\bigskip

As anyone who has read the science columns of the newspapers over the past decade knows, the universe is now accelerating.  The most recent WMAP observations of the Cosmic Microwave Background Radiation provide the strongest evidence for acceleration, but there are several independent lines of evidence that lead to the conclusion that the universe is accelerating.  The evidence is also strong that the mechanism for the acceleration is due to a positive cosmological constant.  If this acceleration were to continue forever, then as Barrow and I showed in our book (Barrow and Tipler, 1986), intelligent life will eventually die out, and the entire theory, which I described in section III, would be false.  If intelligent life is to continue until the very end of time --- as it must if the laws of physics are to hold at all times --- then the universe must eventually stop accelerating, slow down until the expansion stops, and then recollapse to a final singularity.  In this section, I shall outline a mechanism which can cancel the acceleration.  My proposal assumes the validity of the Standard Model of particle physics, a theory which is so far supported by all experiments conducted to date, and which provides only one mechanism for a universal acceleration.

The latest WMAP observations of the Cosmic Microwave Background Radiation (CMBR) have provided the following facts.  First, the universe is 13.7 billion years old.  Second, in the present epoch, the density parameters of the curvature, the ordinary matter, the dark matter, and the dark energy are respectively $\Omega_k << 0.01$, $\Omega_m = 0.04$, $\Omega_{DM} = 0.23$, and $\Omega_\Lambda = 0.73$.  Notice that the subscript on the dark energy is $\Lambda$.  I use this subscript to emphasize that the WMAP data indicate the dark energy looks observationally like the effect of a positive cosmological constant, traditionally written $\Lambda$.  Any correct cosmological theory must be consistent with these observations.

The Standard Model, minimally coupled to gravity, necessarily has a positive cosmological constant.  I predicted in my book (Tipler, 1994) that this cosmological constant would cause the universe to undergo an acceleration.  I argued that this acceleration would occur in the collapsing phase of universal history.  I did not realize that an acceleration could also occur in the expanding phase.  Though I should have, since the Standard Model requires such an acceleration.

The Standard Model requires a positive cosmological constant to cancel the effect of the Higgs vacuum.  Recall that according to the Standard Model, the universe is permeated with a non-zero value of the Higgs field, and it is this non-zero value that breaks the electroweak symmetry and gives mass to all the particles.  But this symmetry breaking is accomplished via the Higgs potential, which for constant Higgs field, acts exactly a very strong negative cosmological constant.  Initially, at the Big Bang singularity, the Higgs field, and hence the Higgs potential, was zero.  But zero is not the lowest value of the potential, so as the universe expanded, the Higgs potential dropped to its lowest value, corresponding to a negative cosmological constant.  Now in special relativity, this negative constant can be re-normalized out of existence.  Not so in general relativity.  Any constant in the matter Lagrangian multiples the invariant volume element, and is equivalent to putting in a cosmological constant in the Lagrangian (Weinberg, 1988).

The value of the negative cosmological constant corresponding to the Higgs potential can be set by experiment, and it is enormous: $ - 1.0 \times 10^{26}$ gm/cm$^3$, as compared to the energy density of the dark matter and dark energy, only $10^{-29}$ gm/cm$^3$.  The only way to make the Standard Model consistent with general relativity is to add a positive cosmological constant of the same magnitude to the Lagrangian.  We would expect the value of the added positive cosmological constant to precisely cancel the value of the Higgs potential, when the Higgs is in its true ground state (the absolute lowest energy density of the potential).

But the Higgs field cannot presently be in its true ground state, for a very simple reason: there is more matter than antimatter in the universe.  The Standard Model has a mechanism of generating this observed excess of matter over antimatter, but most cosmologists believe that this cannot be the main mechanism to generate matter, because they think, incorrectly, that it will generate too many photons to baryons.  I have shown that this large number of photons to baryons is a consequence of imposing the wrong boundary conditions in the very early universe.  If the only boundary conditions consistent with the Bekenstein Bound (a.k.a. quantum field theory) are imposed, the photon to baryon ratio turns out fine.  The Standard Model generation of matter works by electroweak vacuum tunneling.  And if this tunneling yields an excess of matter over antimatter, the Higgs field cannot be in its true vacuum.  Thus the excess of matter over antimatter in the universe ultimately causes the observed acceleration of the universe!

Conversely, if the excess of matter over antimatter were to disappear --- if matter were converted into energy via electroweak tunneling --- and if this disappearance were to occur rapidly enough, then the Higgs potential would fall toward its true ground state, the positive cosmological constant would be progressively cancelled, and the universe would cease to accelerate.  If he universe were a spatially a three-sphere --- and I have argued in the previous section that it is --- then once the acceleration stops, the universe will expand to a maximum size, and then recollapse into the final singularity.

Provided, of course, than a mechanism can be found to convert matter into energy via electrweak quantum tunneling.  The mechanism would have to be the inverse of the process that created the matter excess in the early universe.  But a large amount of matter was created in the early universe because the gauge field energy density was enormous.  The gauge field energy density is tiny today: $10^{-31}$ gm/cm$^3$, and getting smaller as the universe expands.  If the acceleration is to stop, another mechanism must annihilate the matter.

I claim that our future descendants will annihilate the matter.  Once again, they will annihilate the matter in order to survive.  Survival requires energy.  If baryon number is conserved, then only a small fraction of the energy content of matter can be extracted.  If hydrogen is converted into helium, as in the Sun, only 0.7\% of the mass of the hydrogen is converted into energy.  But if our descendants use the inverse of baryogenesis (the technical term for the process that generated matter in the early universe), ALL the energy in matter can be extracted.  I predict that in the future, a way will be found to use inverse baryogenesis, our descendants will use this process as their main energy source, and as a consequence of using up there matter resources, they will save both themselves, and the entire universe.  Because if the acceleration can be cancelled and universal recollapse induced, then the gravitational collapse energy can provide an unlimited energy source, as I showed above.

But in an accelerating universe, life can only travel to the cosmological event horizon, which is about 10 billion light years away at the present time, given the observed value of the dark energy.  (Actually, I should call it the ``pseudo event horizon'', since it would be a true event horizon only if life never stops the expansion, and the Omega Point never develops.  The Omega Point, recall, means that there are no event horizons.)  But quantum non-locality means that the quantum tunneling responsible for baryogenesis generates a uniform density of baryons on large scales.  (And since it is the creation of baryons that generate perturbations in the CMBR, the perturbation spectrum must be scale invariant.)  This means that the baryons have essentially the same density on large scales everywhere in the universe.  This means that the acceleration must be universal.  This means that if the universe is to recollapse, the baryons must be annihilated everywhere, even at distances greater than 10 billion light years, where our descendants cannot travel, even were rockets based on baryon annihilation to be constructed.  Such rockets could approach light speed.   I have shown (Tipler, 1994) that such rockets can travel cosmological distances, using the expansion of the universe itself to slow down the rocket.   Our descendants can reach the pseudo event horizon but no farther.

Thus the laws of physics require there to exist other intelligent species in the universe.  Because of the Limited Resources Argument, the different intelligent life forms must be rare, roughly one species per Hubble volume.  The nearest other intelligent life form must be roughly 10 billion light years away.  But were we to look for them, we would not see them, because at 10 billion light years, we would see their galaxy as it was 10 billion years ago, probably long before their planetary system formed.

\bigskip
\centerline{\bf{VI. Conclusion and Proposed Experiments}} 
\bigskip

But sufficiently advanced radio telescopes MIGHT be able to detect their future presence.  In other words, I shall now argue that there is a role for SETI!  If we cannot detect alien civilizations, we might be able to detect the one-cell organisms out of which they will eventually evolve.  Provided that these organisms already existed 10 billion years ago.

There is some evidence that the one-cell organism that were our own ancestors were around billions of years before the Earth formed 4.6 billion years ago.  William Schopf (1999, p. 77) has discovered structures in the 3,465 $\pm$ 5 million-year-old Apex chert of Australia that closely resemble modern cyanobacteria.   Schopf identified these structures as fossil cyanobacteria, an identification that has been recently challenged.  But I shall assume that his identification is correct, so I can consider the consequences.  

Now cyanobacteria are actually very sophisticated biochemical machines.  If the fossil found by Schopf are indeed cyanobacteria, then all the machinery of prokaryotes, including photosynthetic ability, must have been present on Earth almost as soon as the Earth became capable of sustaining life, about 3.8 billion years ago.  Schopf himself remarks (1999, p. 98) that it seems extraordinary to suppose that this much sophistication could have evolved in the geologically short period between the solidification of the Earth and the date of the Apex fossils.  I agree with Schopf.  If indeed the Apex structures are fossils of cyanobacteria, then these organisms cannot have evolved on Earth.  They must have evolved their observed level of sophistication on some other planet whose star long ago left the main sequence, and in the process, scattered the cyanobacteria throughout interstellar space.

At the Windsor Castle conference, Paul Davies emphasized the consensus opinion that cyanobacteria could survive a trip from one of Solar System's planets, but because of the amount of radiation that they would receive, they could not survive an interstellar journey.  But the evidence Paul cited was theoretical, rather than experimental.  Cyanobacteria are capable of surviving nuclear explosions, and they have been known to live inside nuclear reactors (Schopf, 1999, pp. 232-234).  Given the ability of cyanobacteria to survive radiation, their biochemical complexity, and the evidence that they appeared almost instantaneously on Earth, I think that the preponderance of evidence says that cyanobacteria evolved billions of years before the Earth formed, on a star that has long since disappeared.

This hypothesis has consequences.  First, our interplanetary space probes should find cyanobacteria wherever in the Solar System there is, or has been, liquid water.  But if cyanobacteria have indeed been dispersed throughout interstellar space billions of years before the Earth formed, we would expect to find cyanobacteria, with the same DNA codons and cellular machinery, wherever there is liquid water in the entire Galaxy.  This hypothesis can be rigorously tested only with interstellar space probes.  Incidentally, notice that I've given in passing yet another reason why interstellar probes will eventually be sent out by any intelligent species: to check how related life is in the Galaxy.

But if photosynthetic organisms have existed for billions of years before the Earth formed --- for the order of 10 billion years --- and if our evolution is typical, we would expect intelligent life near the pseudo event horizon to have evolved from organisms, some of which have photosynthetic ability, which existed on liquid water planets 10 billion years ago.  We would also expect there to have been time for the photosynthetic organisms to convert some of these ancient planets' atmospheres into oxygen atmospheres.  This is what we should search for in distant galaxies: the spectral lines of free oxygen.  It has long been known that the oxygen in Earth's atmosphere can be seen at a distance of 10 light years by a one meter orbiting telescope.  A million-kilometer telescope would be able to see free oxygen lines in planetary atmospheres near the pseudo event horizon.  From the arguments above, some such atmospheres must exist.

A million-kilometer telescope is not going to be built in the immediate future.  In the short run, I would propose testing the hypothesis that the excess of matter over antimatter is responsible for the universal acceleration, and that a special boundary condition on the fields of the Standard Model generate the excess of matter over antimatter.  This can be done rather easily, using a modification of the original equipment that discovered the CMBR.  I have shown in (Tipler, 2001, 2005) that if Standard Model physics is responsible for both the dark matter and the dark energy, then the CMBR should not couple to right-handed electrons, and this can be seen by sending the CMBR through filters consisting of poor conductors.  Through such a filter, the CMBR would be more penetrating than thermal radiation of the same temperature.  I have shown elsewhere that the same effect is visible in the Sunyaev-Zel-dovich effect (Tipler, 2005), and it is responsible for the great penetrating power of ultrahigh energy cosmic rays (Tipler, 2001, 2005). 

Two of the arguments against the existence of ETI have been around for a long time.  The evolutionary argument goes back to Alfred Wallace, with Darwin the co-discoverer of the principle of natural selection.  The Fermi Paradox goes back to Enrico Fermi.  I've added a third, the ``Limited Resources Argument'' which connects the rarity of intelligent life in the universe to the unlimited survival of intelligence in the far future.  But to appreciate the power of this argument, we must learn to give up anthropocentric ways of thinking.  We must abandon the (usually tacit) idea that our technology exhausts what is possible using the known laws of physics.  We must abandon the idea that the universe acts according to human thought patterns, that causality works from past to future.  We must abandon the idea that the universe evolves us as the highest level of intelligence, and that all other intelligent species will be as limited in space as we are.  Finally, we must abandon the idea that there is a limit to what intelligence can accomplish, and that intelligence will never play a role on the cosmological scale.  Once we give up these human ways of thinking, we can appreciate the true relation between intelligent life and the cosmos.

\bigskip
\centerline{\bf{References}} 
\bigskip
\noindent
Barrow, J.D., Tipler, F. J. 1986 The Anthropic Cosmological Principle, Oxford University Press.
\medskip
\noindent
Hawking, S.W., Ellis, G.F.R. 1973 The Large-Scale Structure of Space-Time, Cambridge University Press.
\medskip
\noindent
Schopf, W. 1999 Cradle of Life: the Discovery of EarthÕs Earliest Fossils, Princeton University Press.
\medskip
\noindent
Tipler, F. J. 1994 The Physics of Immortality, Doubleday.
\medskip
\noindent
Tipler, F. J., Graber, J., McGinley, M., Nichols-Barrer, J., Staecker 2000 gr-qc/0003082.
\medskip
\noindent
Tipler, F.J. 2001 astro-ph/0111520.
\medskip
\noindent
Tipler, F. J. 2005, Reports Prog. Phys. {\bf 68}, pp. 897--964.
\medskip
\noindent
Weinberg, S. 1989, Rev. Mod. Phys., {\bf 61}, pp. 1--22.

%\vfill\eject

\vfill\eject
 \bye